\documentclass{desyproc}

\usepackage{xspace}

\def\cref#1{Chapt.\,\ref{#1}}
\def\Cref#1{Chapter~\ref{#1}}

\def\fref#1{Fig.\,\ref{#1}}

\def\rref#1{Ref.\,\cite{#1}}

\def\1{\footnotemark[1]}
\def\and{\& }

\def\gcm2{g/cm$^2$\xspace}

\begin{document}
\title{Test of hadronic interaction models with air shower data
\footnote{Invited talk at the 13th International Conference on Elastic \&
Diffractive Scattering, CERN, 2009}
}

\author{{\slshape J\"org R. H\"orandel}\\[1ex]
Radboud University Nijmegen, P.O. Box 9010, 6500 GL Nijmegen, The Netherlands;\\
http://particle.astro.kun.nl}

\contribID{smith\_joe}

\desyproc{DESY-PROC-2009-xx}
\acronym{EDS'09} 
\doi  

\maketitle

\begin{abstract}
The description of high-energy hadronic interactions plays an important role
in the (astrophysical) interpretation of air shower data.  The parameter space
important for the development of air showers (energy and kinematical range)
extends well beyond todays accelerator capabilities.  Therefore, accurate
measurements of air showers are used to constrain modern models to describe
high-energy hadronic interactions.  The results obtained are complementary to
information gained at accelerators and add to our understanding of high-energy
hadronic interactions.
\end{abstract}

\section{Introduction}

The understanding and modeling of extensive air showers (particle cascades in
the atmosphere) brings together the particle physics and astroparticle physics
communities. To strengthen the connections and the scientific exchange between
those communities is very fruitful for both sides and yields complementary
information on the understanding of high-energy hadronic interactions.

When high-energy cosmic rays impinge onto the atmosphere they initiate cascades
of secondary particles -- the extensive air showers.  Observations of air
showers are used to indirectly infer the properties of cosmic rays at energies
exceeding $10^{14}$~eV.  The interpretation of air shower data faces a twofold
challenge: the (exact) mass composition of cosmic rays is not known at those
energies and, additionally, the properties of high-energy interactions taking
place in air showers are partly unknown.  Direct measurements of cosmic rays
(fully ionized atomic nuclei) at energies below $10^{14}$~eV indicate that they
are mostly composed of elements from hydrogen (protons) up to iron
\cite{cospar06,behreview}. The abundance of heavier elements is significantly
smaller. Hence, in the following, we assume that cosmic rays comprise elements
from hydrogen to iron.

We will focus on results from the KASCADE experiment \cite{kascadenim}, one of
the most advanced air shower detectors in the energy range around $10^{15}$~eV.
It has a unique set-up which allows to measure simultaneously the
electromagnetic, muonic, and hadronic shower components.  This is in particular
valuable to test the consistency of hadronic interaction models.  Since about a
decade \cite{Horandel:1998br,wwtestjpg} systematic checks of interaction models
are performed with air shower data and the most stringent constraints on
interaction models, derived from air shower data have been obtained with
KASCADE measurements.

KASCADE consists of several detector systems. A $200 \times 200$~m$^2$ array of
252 detector stations, equipped with scintillation counters, measures the
electromagnetic and, below a lead/iron shielding, the muonic parts of air
showers.  An iron sampling calorimeter of $16 \times 20$~m$^2$ area detects
hadronic particles \cite{kalonim}. It has been calibrated with a test beam at
the SPS at CERN up to 350~GeV particle energy \cite{kalocern}.  For a detailed
description of the reconstruction algorithms see \rref{kascadelateral}.

\section{Quantitative tests}

The principle idea of the consistency tests of hadronic interaction models is
to simulate air showers initiated by protons and iron nuclei as the two
extremes of possible primary particles.  The shower simulations were performed
using CORSIKA \cite{corsika}, applying different embedded hadronic interaction
models.  In order to determine the signals in the individual detectors, all
secondary particles at ground level are passed through a detector simulation
program using the GEANT package \cite{geant}.  
\footnote{ For details on the event selection and reconstruction, see
\rref{jensjpg,epostest,atteas}.}
The predicted observables at ground level, such as e.g.\ the number of
electrons, muons, and hadrons or the energy of the hadrons are then compared to
the measurements. If the measured values are inside the predicted interval for
proton and iron induced showers, the particular interaction model used for the
simulations is compatible with the data. On the other hand, if the measured
values are outside the proton-iron interval, there is a clear hint for an
incompatibility between the model under investigation and the measurements.

Hadronic interactions at low energies ($E_h<80$ and 200~GeV, respectively) were
modeled using the {\bf GHEISHA} \cite{gheisha} and {\bf FLUKA}
\cite{flukacern,flukaCHEN} codes.  Both models are found to describe the data
equally well \cite{jensjpg}.  High-energy interactions were treated with
different models as discussed below, several models have been systematically
tested over the last decade.  

\begin{figure}
 \includegraphics[width=0.49\textwidth]{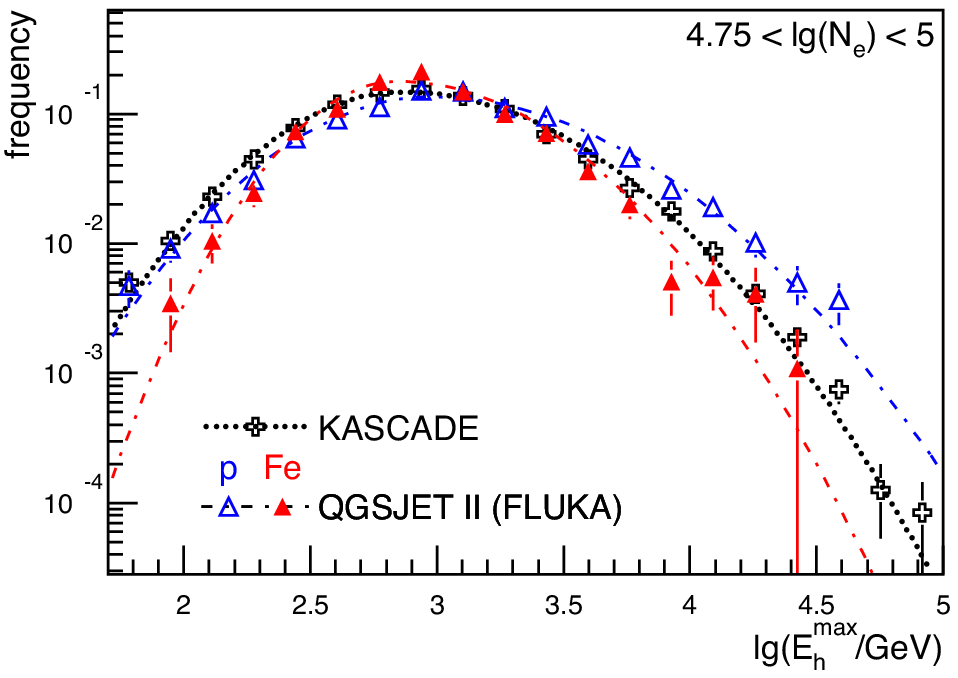}\hspace{\fill}
 \includegraphics[width=0.49\textwidth]{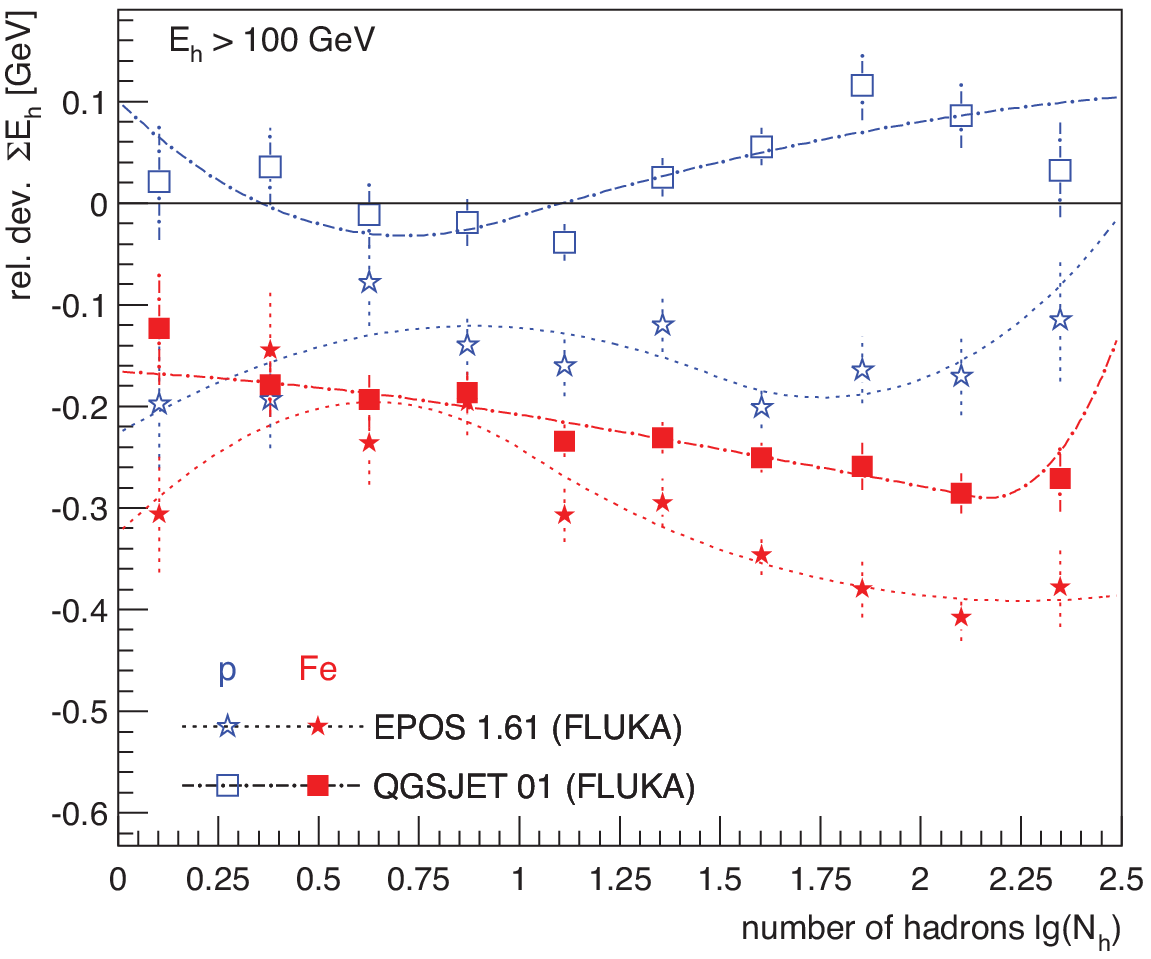}
 \begin{minipage}[t]{0.49\textwidth}
  \caption{Energy of the most energetic hadron reconstructed at
   observation level.  Predictions of QGSJET~II are compared to measured values
   \cite{icrc09-hoerandel}.\label{qgs}}  
 \end{minipage}\hspace{\fill}
 \begin{minipage}[t]{0.49\textwidth}
  \caption{Relative hadronic energy sum $(\sum
   E_h^{sim}-\sum E_h^{meas})/\sum E_h^{meas}$ as function of the reconstructed
   number of hadrons for two interaction models and two primary particle species
   \cite{epostest}.\label{epos}}       
 \end{minipage}
\end{figure}

First quantitative tests \cite{Horandel:1998br,wwicrc99,wwtestjpg} established
{\bf QGSJET}~98 \cite{qgsjet} as the most compatible code. Similar conclusions
have been drawn for the successor code QGSJET~01 \cite{jensjpg}.  The next
version of the code, QGSJET-II-2, has been investigated recently
\cite{icrc09-hoerandel}. The analyses yield inconsistencies, in particular for
hadron-electron correlations. An example is shown in \fref{qgs}.
For a given interval of the number of electrons the frequency of the maximum
hadron energy registered in each shower is plotted.  It can be recognized that
for small hadron energies, the measured values are outside the range
(proton-iron) as predicted by QGSJET-II-2.  Studies of the latest version,
QGSJET-II-3, are in progress.

Predictions of {\bf SIBYLL}~1.6 \cite{sibyll16} were not compatible with air
shower data, in particular there were strong inconsistencies for hadron-muon
correlations \cite{Horandel:1998br}. These findings stimulated the development
of SIBYLL~2.1 \cite{sibyll21}. This model proved to be very successful, the
predictions of this code are fully compatible with KASCADE air shower data
\cite{jenskrakow,jenspune,jensjpg}. 

Analyses of the predictions of the {\bf DPMJET} model yield significant
problems in particular for hadron-muon correlations for the version DPMJET~2.5
\cite{dpmjet}, while the newer version DPMJET~2.55 is found to be compatible
with air shower data \cite{jensjpg}.

Investigations of the {\bf VENUS} \cite{venus} model revealed some
inconsistencies in hadron-electron correlations \cite{wwtestjpg}.  The
predictions of {\bf NEXUS}~2 \cite{nexus} were found to be incompatible with
the KASCADE data, in particular, when hadron-electron correlations have been
investigated \cite{jensjpg}.

Recently, predictions of the interaction model {\bf EPOS}~1.61
\cite{epos,eposmerida,epos2} have been compared to KASCADE air shower data
\cite{epostest}.  This model is a recent development, historically emerging
from the VENUS and NEXUS codes.  The analysis indicates that EPOS~1.61 delivers
not enough hadronic energy to the observation level and the energy per hadron
seems to be too small. This is illustrated in \fref{epos}: the
predicted hadronic energy sum, relative to the measured values is plotted as
function of the number of reconstructed hadrons. In this representation the
measured values are at the zero line.  Shown are results for two interaction
models and two primary particle species. The values for protons and iron nuclei
for QGSJET~01 are above and below zero, respectively, as expected. However, for
EPOS the predictions for both primary particle types are significantly below
zero. A strong hint that the predictions are not compatible with the data.
Most likely, the incompatibility of the EPOS predictions with the KASCADE
measurements is caused by too high inelastic cross sections for hadronic
interactions implemented in the EPOS code.  These findings stimulated the
development of a new version EPOS~1.9 \cite{icrc09-pierog}.  Corresponding
investigations with this new version are under way.

Presently, the most compatible predictions are obtained from the models
QGSJET~01 and SIBYLL~2.1.

For a more detailed test of the interaction models one has to assume a mass
composition in the simulations to compare a single simulation curve (instead of
a proton-iron range) with the measured distribution. This can be done
consistently by taking a mass composition derived from other observables using
the same combination of low-energy and high-energy models \cite{jenspune}.
Energy spectra for elemental groups in cosmic rays have been obtained by
applying an unfolding procedure to the measured two-dimensional electron and
muon number spectra \cite{ulrichapp}.  This composition of cosmic rays has been
used as input for the air shower simulations and the predicted observables at
ground level have been compared to the measurements.  The investigations reveal
that the deviations between the model predictions and the measurements are of
the order of 15\% \cite{jenspune}.  This number illustrates the present
accuracy of the quantitative description of the development of air showers.

At much higher energies (exceeding $10^{18}$~eV) investigations of
hadronic interactions are under way with data from the Pierre Auger
Observatory. At present, all models investigated exhibit problems in predicting
the correct number of muons in air showers
\cite{engelmerida,Schmidt:2009ge,icrc09-castellina,Ulrich:2009hc}. 

\section{Uncertainties of accelerator measurements extrapolated to air shower
observables}

Also studies have been carried out to evaluate the effect of uncertainties in
the description of individual interactions on the development of air showers.
An example is the variation of the inelastic proton-proton cross section and
the elasticity of the interactions  within the error bounds given by
accelerator measurements \cite{wq}.  For the studies parameters in the hadronic
interaction model QGSJET~01 have been modified.
For illustration, we use here the models 3 and 3a from \rref{wq}.
\footnote{Model 3 and 3a refer to nomenclature
used in \rref{wq}. For model~3 the cross section has been lowerd, and, in addition for model~3a the elasticity has been modified.} 
With respect to the original QGSJET~01 code
the inelastic hadronic
cross sections have been lowered, e.g.\ the proton-air cross section at
$10^6$~GeV is reduced by 5\% from 385~mb to 364~mb and the elasticity has been
increased by about 12\%. 

A lower cross section implies a longer mean free path for the hadrons in the
atmosphere and thus a reduction of the number of interactions. A larger
elasticity means that more energy is transferred to the leading particle.  Both
changes applied result in showers which penetrate deeper into the atmosphere.
For example, the average depth of the shower maximum for protons at 100~PeV is
shifted by 24~\gcm2 due to the lower cross section and by 10~\gcm2 due to the
higher elasticity \cite{wq}.

The shift of the shower maximum also affects the number of particles registered
at ground level.  Since the maximum moves closer to the observation level one
expects an increase of the number of particles.  However, reducing the number
of interactions due to a lower cross section also reduces the possibility to
produce secondary particles and an increase of the elasticity implies at the
same time that less energy is available for multi-particle production.  This
means that we are faced with two competing processes influencing the number of
particles observed.

\begin{figure}
 \includegraphics[width=0.49\textwidth]{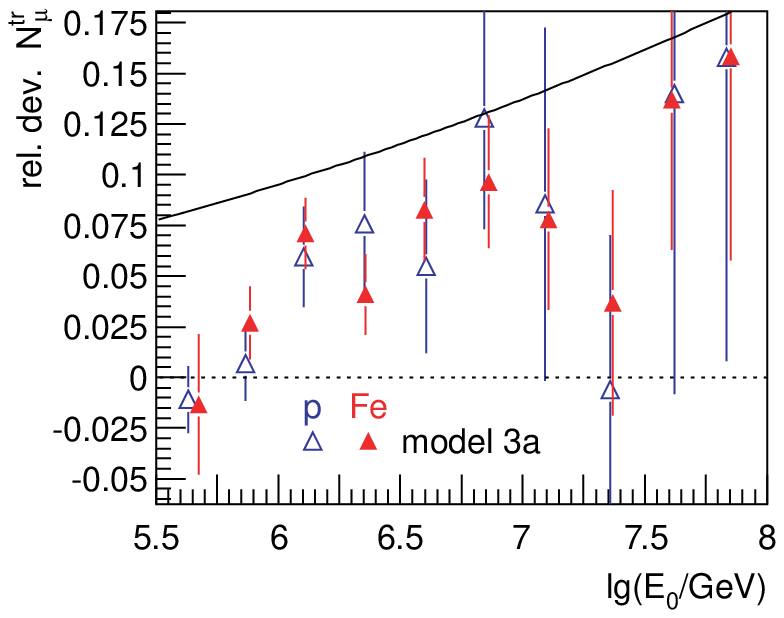}\hspace{\fill}
 \includegraphics[width=0.49\textwidth]{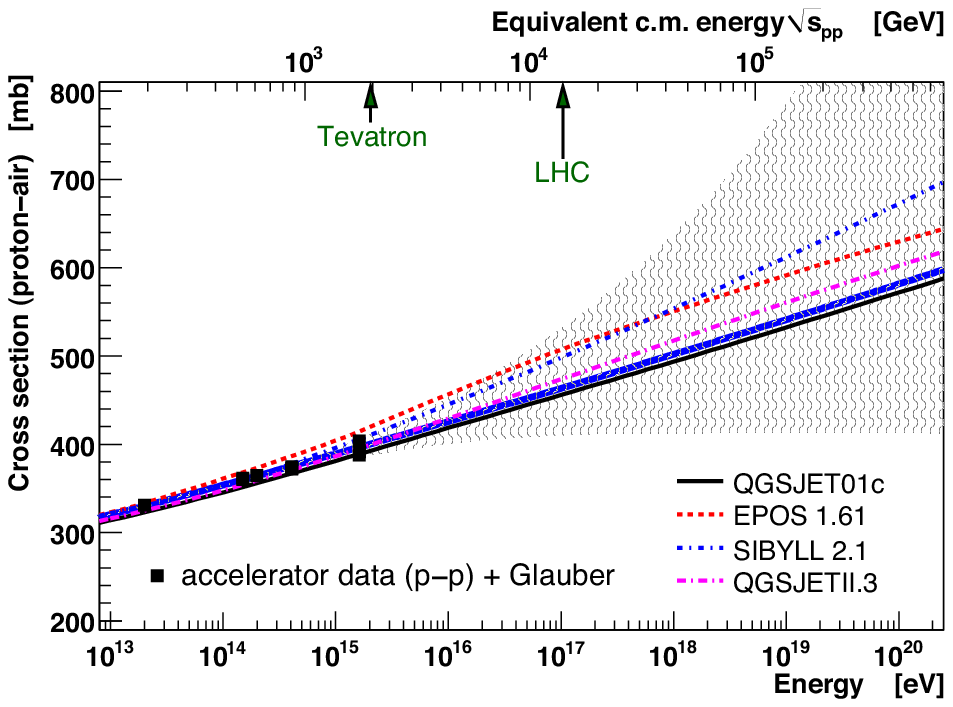}
 \begin{minipage}[t]{0.49\textwidth}
  \caption{Relative deviation of the number of muons in model~3a
   relative to model~3, i.e. the change of the number of muons related to an
   increase of the elasticity as function of primary energy. The line indicates
   an estimate according to a simple Heitler model
   \cite{kascadewqpune}.\label{nme0}}
  \end{minipage}\hspace{\fill}
  \begin{minipage}[t]{0.49\textwidth}
   \caption{Uncertainties of the extrapolation of the proton-air cross
    section from accelerator to cosmic-ray energies
    \cite{Ulrich:2009yq}.\label{ppextra}}
  \end{minipage}
\end{figure}

Simulations reveal that an increase of the elasticity enhances the particle
numbers for all species observed (electrons, muons, and hadrons). An increase
is registered for both, primary protons and iron nuclei. This means the effect
of deeper penetrating cascades seems to dominate.  As an example, the increase
of the number of muons when increasing the elasticity is illustrated in
\fref{nme0} \cite{kascadewqpune}. Shown are the relative changes in the
number of muons for model~3a 
relative to model~3 ($\delta
N_\mu=(N_\mu^{3a}-N_\mu^{3})/N_\mu^{3}$) for primary protons and iron induced
showers as function of primary energy.

The increase of the number of muons $N_\mu$ as function of primary energy $E_0$
has been estimated using a Heitler model to be $N_\mu=(E_0/\xi_c^\pi)^\beta$,
where $\xi_c^\pi\approx20~GeV$ is the critical energy for pions at which the
probability for an interaction and decay are about equal
\cite{matthewsheitler}. The exponent $\beta$ depends on the elasticity of the
interaction as $\beta\approx1-0.14(1-\kappa)$. Using the energy dependence of
$\kappa$ for the two modifications of QGSJET \cite{wq} and introducing an
energy dependent $\beta$, an increase of the number of muons as function of
energy is expected as indicated by the line in \fref{nme0}. The
general trend of the simple estimate is reflected by the detailed simulations,
but the absolute values are about 5\% larger for the simple estimate as
compared to the full simulation. This illustrates the sensitivity of air shower
observables to properties of hadronic interactions. Another example is
discussed in the following.

Recent investigations \cite{atteas} revealed that the attenuation length
$\lambda_E$, defined as $ \Sigma E_H=E_0\exp\left(-X/\lambda_E\right)  $ is
very sensitive to the inelastic hadronic cross sections. $E_0$ is the energy of
the primary particle and $\sum E_H$ the hadronic energy sum registered at
ground level. Thus, $\lambda$ is a measure for the hadronic energy transported
to ground.
A detailed inspection of the attenuation lengths obtained for showers induced by
light and heavy elements indicates that the cross sections in the
hadronic interaction model QGSJET\,01 may be too large and the elasticity may
be too small. A modification with altered parameters (model~3a as discussed
above) improves the situation.

Also the influence of the transverse momentum $p_\perp$ in hadronic
interactions on (hadronic) air shower observables has been analyzed by the
KASCADE group \cite{annaprd}. It turned out that the geometrical distributions
of the most energetic hadrons at ground level are sensitive to this parameter.
The maximum geometrical distance $d_{max}$ between the four highest-energy
hadrons in each shower are sensitive to $p_\perp$.  Altering $p_\perp$ in air
shower simulations results in different $d_{max}$ distributions.

\section{Accelerator data needed for cosmic-ray physics}

Complementary to the investigation of air showers
more information about hadronic interactions is needed from accelerator
experiments to fully understand cosmic rays, as discussed in the following.

\paragraph{Air shower measurements}
In high-energy interactions most energy is escaping the interaction region in
the forward direction, i.e.\ at large pseudorapidity values $\eta$.  For
example, the energy flow at the LHC at $E_{cm}=14$~TeV, corresponding to
$E_{lab}\approx2 E_{cm}^2/m_p=10^{17}$~eV, peaks at pseudorapidity values around 7 to
10. The forward region with values $|\eta|>4$ is of great importance for air
shower experiments.

Of particular interest are the total (inelastic) cross sections, the
elasticity/inelasticity of the interactions, as well as the production cross
sections of secondary particles and their parameter distributions, like
multiplicity, transverse momentum, energy, and pseudorapidity.  As projectiles
protons and pions are of interest to study the elementary interactions but also
beams of heavier nuclei (such as C, N, O, or Fe, being dominant in the
cosmic-ray composition) are desirable.  Targets are preferably air
constituents, i.e.\ nitrogen, oxygen, (and carbon).  In particular, at the LHC
the study of p-p  and p-N interactions is of great importance.

The uncertainties introduced in the proton air cross section by extrapolating
from accelerator data to highest energies is illustrated in \fref{ppextra}
\cite{Ulrich:2009yq}. It is obvious that LHC data will drastically
reduce the uncertainties in the regime of the highest-energy cosmic rays.

\paragraph{Direct measurements} 
Further input from accelerator experiments is also required for the
interpretation of data from balloon borne cosmic-ray detectors, delivering
unique information about the propagation of cosmic rays in our Galaxy. The
systematic uncertainties of measurements of the boron-to-carbon ratio are
presently dominated by uncertainties in the production cross section of boron
in the residual atmosphere above the detector \cite{creambc}. Boron is produced
through spallation of the relatively abundant elements of the CNO group in the
atmosphere.
\footnote{The detectors float typically below a residual atmosphere of about
$3-5$~\gcm2.}
Thus, the production cross sections of boron for protons and CNO nuclei
impinging on nitrogen targets are of great interest at energies significantly
exceeding 100~GeV/n.

\section{Outlook}
We are looking forward to new data from the LHC in the next few years. They
will improve current models used for air shower simulations. 
Complementary, at energies much higher than at the LHC, data from the Pierre
Auger Observatory will yield further information on high-energy hadronic
interactions.
A close cooperation between the high-energy physics and
astroparticle physics communities will help to improve our understanding of
elementary processes in nature.

\section*{Acknowledgments}
It was a pleasure for me to serve together with Gianni Navarra as convener for
the cosmic-ray session at the EDS 09 meeting. I thoroughly regret that Gianni
Navarra passed away in August 2009 --- we miss him and will always remember
him.

I would like to thank the organizers of EDS~09 for the possibility to
strengthen the connection between the accelerator and cosmic-ray communities.
In particular, I acknowledge the great efforts of Mario Deile to organize a
perfect meeting.

I'm grateful to my colleagues from the KASCADE-Grande experiment for fruitful
discussions.


\begin{footnotesize}



\end{footnotesize}


\end{document}